# I'm all Ears! Listening to Software Developers on Putting GDPR Principles into Software Development Practice

*Abdulrahman Alhazmi*

*School of Engineering and Mathematical Sciences at La Trobe University*

20509556@students.latrobe.edu.au

*Nalin Asanka Gamagedara Arachchilage*

*Optus La Trobe Cyber Security Research Hub*

*School of Engineering and Mathematical Sciences at La Trobe University*

nalin.asanka@latrobe.edu.au

**ABSTRACT**

*Previous research has been carried out to identify the impediments that prevent developers from incorporating privacy protocols into software applications. No research has been carried out to find out why developers are not able to develop systems that preserve-privacy while specifically considering the General Data Protection Regulation principles (GDPR principles). Consequently, this paper aims to examine the issues, which prevent developers from creating applications, which consider and include GDPR principles into their software systems. From our research findings, we identified the lack of familiarity with GDPR principles by developers as one of the obstacles that prevent GDPR onboarding. Those who were familiar with the principles did not have the requisite knowledge about the principles including their techniques. Developers focused on functional than on privacy requirements. Unavailability of resourceful online tools and lack of support from institutions and clients were also identified as issues inimical to the onboarding of GDPR principles.*

Keywords: Usable security, Privacy, Software developers, Secure software engineering, GDPR.

1. **INTRODUCTION**

The use of software applications on a wider scale has resulted in clients disseminating their information more frequently and widely [1]. The data is used and disseminated using methods that are difficult for clients to comprehend [2]. This could be because applications are developed without including privacy protocols such as General Data Protection Regulation (GDPR) principles. The GDPR principles clarify six principles [3] that give developers some useful guidelines. The use of these guidelines ensures the development of software systems that are capable of safeguarding privacy [4]. Data privacy is in the top five IT management issues and is also listed in the top five biggest IT investments [30]. Losses from data breaches have always been a concern. Large funds have been set aside to secure information systems and deal with security lapses in organizations [31] [39]. Many activities have been done to combat data breaches [40]; however, they still occur, as evidenced by the recent Zoom data breach [32]. Zoom experienced a boom during the Covid-19 pandemic as many meetings were held remotely so as to keep social distance. Many users imply higher revenue, however due to privacy issue, a significant percentage of the users stopped using zoom [34]. According to [34], a study that was done to investigate how different professionals have been affected by zoom security and privacy issue, it was found that 12.1% of professionals completely stopped using Zoom, which implies reduction of revenue. Another example is the Cambridge Analytica scandal [36]. The Cambridge Analytica-Facebook scandal contributed to public concern about the techniques used by Cambridge Analytica to manipulate voters, based on Facebook user data, through psychographic profiling algorithms. The scandal eventually led to the Federal Trade Commission (FTC) placing a record-breaking $5 billion penalty on Facebook in July 2019. Software applications, which are developed without the ability to preserve privacy, are likely to be affected by data breaches [4]. Previous research has revealed that software developers may not adhere to these data privacy principles when implementing privacy-preserving software systems [5], [6], [7] [8].

There have been other serious data breaches on application systems that hold loads of clients' data such as Facebook [5], [9], [10], [11], Yahoo [12] and more recently, [13]. The data breaches may indicate that different parties responsible for ensuring privacy are implemented cannot incorporate the necessary privacy protocols into applications they develop despite the existence of laws such as GDPR principles. Data breaches frequently happening on different scales imply that software applications remain unsuccessful in safeguarding clients' privacy due to their inability to enforce GPDR law (i.e. GDPR principles).

Though debatable, the issue of privacy could be dealt with by positively influencing the conduct of developers through the enforcement of important implementation practices. These practices will ensure the incorporation of privacy in application systems. Such developers may also require assistance because they lack specialist knowledge in privacy and security issues [14]. Precious researchers have found out that developers lack resources such as step by step frameworks on privacy implementation and proper knowledge on techniques such as Pseudonymisation, anonymization and encryption. [5], [10], [11], [15]. These techniques translate privacy requirements into software systems and would have enabled the inclusion of privacy in systems if they were readily available to developers. Some researchers e.g. [5] found out that most of his participants had never had of privacy theories, therefore they can't implement privacy techniques not known to them. According to [35], Special privacy protection requirements should be taken into consideration at an early stage, as modern technical technologies frequently involve hidden hazards that are very difficult to resolve once the basic architecture has been worked out. Instead, then trying to come up with laborious and time-consuming fixes, it makes all the more sense to recognize and analyze potential security and privacy issues while designing new systems and to integrate privacy protection into the overall design. Therefore, software designers, who play critical role in privacy by design, must understand privacy requirements which are guided by privacy laws such as GDPR for example personal data should be anonymized according to GDPR, a developer who understands GDPR would implement this. [29] Provided information that if software developers fail, the cyber-security system fails, and this may lead to data breaches. There is, therefore, the need to find out why developers are not able to include GDPR principles when developing privacy-preserving software systems.

This paper aims to explore the problems developers encounter when including privacy concerning all the GDPR principles. Our initial findings revealed that developers lack familiarity with GDPR principles as one of the obstacles that prevent GDPR implementation. Those familiar with the principles did not have the requisite knowledge about the principles, including their techniques. We also found out that developers were more focused on functional than on privacy requirements. Unavailability of resourceful online tools and lack of support from institutions and clients were also identified as issues that prevent GDPR implementation.

2. LITERATURE REVIEW

Researchers have attempted to study some of the principles of the General Data Protection Regulation (GDPR). However, no research has been conducted to explore all of the principles of the GDPR. Research paper [16] presented the development of a scale to measure software developers' attitudes towards how they handle personal data in the software they develop. Their findings identified a model consisting of three factors that allow for understanding developers' attitudes: informed consent, data minimization, and data monetization. From the results, they discuss mismatches between developers' attitudes and their self-perceived extent of properly handling their users' privacy, and the importance of understanding developers' attitudes towards data monetization. However, the authors did not take into account other principles covered by the GDPR, such as purpose limitation, accuracy, storage limitation, integrity, and confidentiality. As noted, it is imperative to investigate when software developers were asked to integrate privacy into the systems they develop. One can argue that to preserve user privacy in software applications, GDPR principles must be complied with [3].

The research paper [17] surveyed six experienced senior engineers with the aim of understanding what motivates them to apply privacy regulations. Their findings revealed a lack of perceived responsibility, control, autonomy, and frustrations with interactions to the legal world. The research paper [18] in their report found that privacy and data protection features are ignored by traditional engineering approaches when implementing the desired functionality. They further found that ignorance is caused and supported by limitations of awareness and understanding of developers and data controllers as well as lacking tools to realize privacy by design. A research article titled "How Developers Make Design Decisions about User Privacy" [19] explored whether specific influences, including organization, perceived professional and personal privacy, affect developers' decision-making about privacy [19]. This research did not investigate what developers face when incorporating GDPR-compliant privacy.

The research paper [5] also carried out an experiment considering 36 software developers in a software design task. They achieved this task by using instructions for integrating privacy to identify the problems they. Their research investigated the principle of data minimization of the GDPR. According to [3], studying the technique of data minimization alone is not enough to solve the issues of software privacy. It could imply the failure to implement the GDPR. The research [5] was also more attentive to the data minimization principle of the GDPR.

The research paper [21] was more concerned with organization privacy practices of organizations. It intensely studied the environmental mechanisms and identified the environmental components that affect and, are affected by developers when dealing with privacy concerns. Additionally, it identified the organizational privacy climate as a good tool that organizations can use to control developers concerning specific translations of privacy. [21] Also studied how users and developers perceive data privacy. They also found users' responses to their privacy concerns. Their surveys show that most users are concerned with their data and location rather than communication data. They also found that people from different regions had different privacy issues. They further discovered that developers mainly focused on anonymization and technical measures. [21] Collected data via an online survey to determine developers' professional approaches and practices concerning privacy. They showed that the developers' decisions on privacy include different characteristics, such as organizational privacy, professional and personal privacy. Likewise, the researchers did not seek to investigate the challenges developers face when implementing privacy and adopting the GDPR.

Another research titled "Us and Them - A Study of Privacy Requirements in North America, Asia, and Europe", discussed the perceptions of developers and users of privacy in detail [22]. It also included the critical aspects of privacy and the best techniques to deal with them. It revealed that the expertise of developers in software development has an impact on obtaining confidentiality requirements [22]. Finally, the article also depicted that geography has an impact on confidentiality requirements [22]. Like previous research, this study did not take into account the reasons why developers do not comply with the GDPR when dealing with privacy in software development.

Despite all the research, no one has identified the challenges that developers face when integrating user data privacy following all of GDPR principles, which are obligatory for all developers, software vendors and organizations to comply [23]. This research aims to empirically study the problems that software developers face when trying to integrate privacy into software systems taking GDPR on board. By this, it will observe those who are asked to embed privacy into the applications they design using GDPR and then relate those issues into guidelines that should be considered when establishing privacy practices for software developers.

3. **METHODOLGY**

This research was designed to involve developers from all over the world regardless of their gender. For this research to be more diverse, developers of different ages and with different years of experience were recruited i.e. novice and experienced. According to [37I novice web programmers, with little or no secure programming skills, unknowingly develop web applications ripe with security vulnerabilities, thus compromising the integrity of the application. Neither a novice nor an experienced programmer necessarily knows how to write code securely [38]. Therefore, it is important to investigate why both novice and experienced cannot implement privacy correctly as this may reduce vulnerably in software systems. The requirement to a participant was that the participant should be a software designer and/or developer.

The primary objective of this research was to identify the challenges that developers face when they try to implement privacy using the GDPR principles. This research took a qualitative research approach [24], which helped in clearly understanding the issues faced by the developers. To identify the issues, we collected and later analyzed qualitative data from the participants using grounded theory-based analysis.

a. **Experimental Setup**

We invited software developers from different countries. We included developers from Europe where GDPR law was founded and those outside Europe. This was aimed at achieving diversity, i.e. whether locality could also have an effect on their knowledge about GDPR. From the participant, only three developers were from Europe. For one to be a participant in this research, the requirement was that he or she had to be a software developer practicing software design and/or architecture. Developers were identified on LinkedIn where users publish detailed proficiency information and participate in interest groups. After identification, an invitation email was sent to the potential participants. Those that agreed to participate marked "I agree" in the participant information consent form which was sent to them. After participants had agreed to participate in this research, they read and understood a scenario and Unified Modeling Language (UML) diagrams, designed to reflect GDPR. This was followed by a cognitive walk-through done remotely via Zoom, which took at most 60 minutes to complete. The cognitive walk-through is a usability evaluation method in which one or more evaluators work through a series of tasks and ask a set of questions from the perspective of the user [25]. Cognitive walk-through evaluates the degree of difficulty to accomplish tasks; thus, this helped to understand the problems faced by the developers [25]. The main reason for choosing this method was that it is less time consuming, hence allowed collection of information for a shorter time. The scenario below was used.

### b. Scenario Development and Procedure

*Participants are required to structure an online health application (web application) that permits patients to register/login and discuss remotely with clinical experts for a fee. Clients should be in a position to communicate with the clinical experts about their issues in this application. The clinical expert can join this system to earn from it. The application will receive a commission from each paid service by the clients.*

In the scenario above, the six principles of GDPR [3] law were investigated and how each principle was tested can be found in table 1. The six principles are:
Lawfulness, Fairness, and Transparency, which states that collected data should be processed lawfully, fairly, and in a transparent way i.e., data subject should be updated on the how their data is used and notified in case of a breach. Purpose Limitation, which prevents those who collect personal data from using it for new purposes if they are incompatible with the original stated purpose to the data subject. Data Minimization, which states that the only data that is required in a given function, will be collected. Accuracy, which states that the collected data from data subjects should be kept accurately and always up to date. Storage Limitation, which states that data should not be held by the system if it is no longer needed. Integrity and Confidentiality: Integrity states that unauthorized persons should not edit data, while Confidentiality states that unauthorized people should not access data. The explanation of how each principle was tested using this scenario can be found in table 1.

This scenario enabled us to research the Accuracy principle. In this health application, one of the requirements was to approve the registration of users, as shown in figure 1 below. This was to ensure that the data collected was a true user record. It was also a requirement in this application to prevent unauthorized persons from accessing users' data as they could tamper with the accuracy of the data. We requested each participant to explain how he or she could achieve this in the health application.

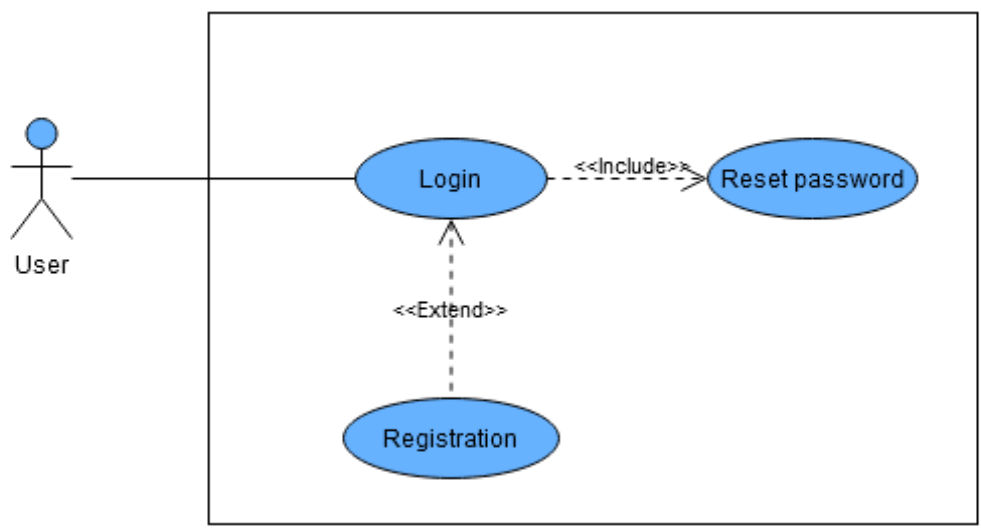

**Fig 1: The registration and login application.**

The second principle we investigated is Integrity and Confidentiality. Integrity states that unauthorized persons should not edit data [3]. This application required data to be protected and prevented from getting into the wrong hands, as this could lead to tampering with the data. The application had a functional requirement that allowed different users to communicate remotely. It was required that security measures are applied such that even if this data gets into the wrong hands, it cannot be edited or read, as shown in the example below in figure 2.

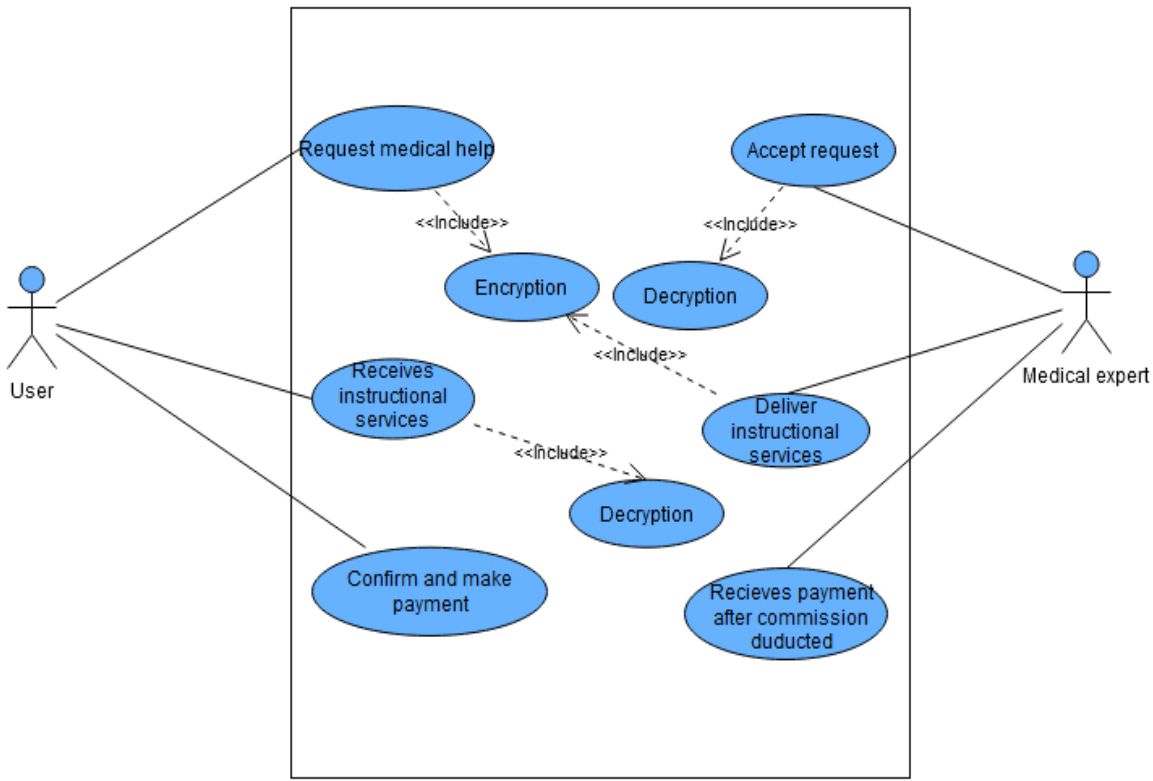

**Fig 2: The use case for the health system.**

We requested developers to state how they can achieve this and the techniques they will use in the given health application. Confidentiality, on the other hand, states that unauthorized people should not access data [3]. In this health application, it was a requirement for users to authenticate before accessing the application. The data storage security measures were also implemented in this application to ensure that unauthenticated persons, do not access data, as shown in the example below in figure 3.

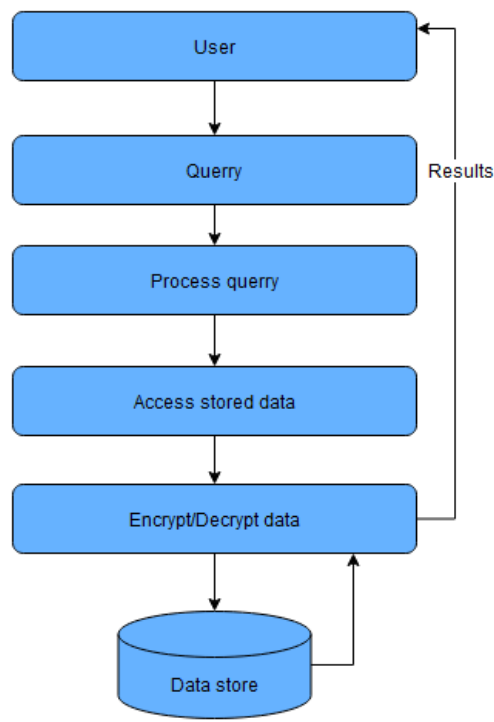

**Fig 3: Information flow to the database.**

Developers taking part in this research were requested to explain how they could achieve that in the health application. This helped in understanding how efficiently they implement this principle and identify any problems they encounter.

Another principle that was investigated with the scenario is Lawfulness, Fairness, and Transparency. In this scenario, we collected doctors, nurses, and clients' data. Developers were required to share up-to-date information with the data subjects on how their data will be processed. They were also required to promptly notify data subjects of any data breaches when they occur. The developers were requested to explain how they would achieve these functions in the health application.

The fourth principle of GDPR we investigated is Data Minimization. In this health scenario, there are functional requirements such as login, registration, and payment process. One of the needs of this application is to collect the least user data as possible in each function. For example, the login may only require an email and password. We asked participants of the research to explain how they can accomplish this need in this health application. The scenario also enabled us to investigate the Purpose Limitation principle. This principle prevents those who collect personal data from using it for new purposes if they are incompatible with the original purpose for collecting the data [3]. In this health application, it is a requirement to ensure that data subjects know the reason for each data that is collected. It will also be required that this data should not be used for any other purposes than the stated ones. Participants of this research were required to explain how they could achieve this in this application. The sixth principle of the GDPR law is Storage Limitation. This health application requires that the stored records should not be held by the system for a long period once they are no longer needed. This will ensure that only minimum data will leak to the unwanted persons if there are data breaches. We were able to investigate this in the scenario, as the participants of this research were requested to explain how they would use different techniques to achieve.

**Table I: Method of testing different principles**

| Principle | Testing Method |
|---|---|
| Lawfulness, Fairness and Transparency | In this scenario, we collected doctors, nurses, and clients' data. Developers were required to share up-to-date information with the data subjects on how their data will be processed. They were also required to promptly notify data subjects of any data breaches when they occur. The developers were requested to explain how they would achieve these functions in the health application. |
| Integrity and Confidentiality | This application will require data to be protected from getting into the wrong hands, as this can lead to tampering with the data. The application also has a functional requirement that allows different users to communicate remotely. It was required that security measures are applied so that even if this data gets into the wrong hands, it cannot be edited or be read. Developers were requested to state how they will achieve this and the techniques they will use. In this health application, it was also a requirement for users to authenticate before accessing the application. Data storage and security measures were also implemented in this application to ensure that unauthenticated persons do not access the data. Developers taking part in this research were requested to explain how they can achieve that in the given health application. |
| Purpose Limitation | In this health application, it was a requirement to ensure that the data subjects know the reason for the collection of each data. It was also required that this data was not to be used for any other purposes other than the stated ones. Participants of this research were required to explain how they could achieve this in this application. |
| Data Minimization | In this health scenario, there were functional requirements such as login, registration, and payment process. One of the needs of the health scenario was to collect the least amount of user data as possible in each function. For example, the login may only require an email and password. Participants of the research were asked to explain how they would accomplish this need in the health application. |
| Accuracy | In this health application, one of the requirements was to approve the registration of users. It was also required to prevent unauthorized persons from accessing users' data as they can tamper with the accuracy of the data. Each participant had to explain how he would achieve this for this health application. |
| Storage Limitation | This health application required that the stored records would not be held by the system for an extended period once they are no longer needed. The participants of this research were requested to explain how they would use different techniques to achieve this. |

### c. Ethics

Our University Ethic Board approved this project. We were requested to download the consent forms, which we sent to the participants for them to provide consent. We were much concerned with the usage and storage of the data collected to maintain the privacy of the participants. Our study did not collect any sensitive information from the participants. We assured the participants that we would maintain confidentiality. In general, we complied with GDPR when dealing with the participants' data. No personal data was published in this investigation. Lastly, we ensured participants that they would be informed of the results of the study.

### d. Final Study

We sent invitation messages to 50 developers. The messages contained the scenario, the questions for the interview, and the consent form. A total of 22 developers agreed to take the interview, and they indicated this by marking, "I agree to start an interview" in the participant consent form. All participants in this study were professional developers practicing software development and/or design. According to stack overflow survey data 2020 [33] which involved 65000 developers, most professional developers are in their or early 30s. Our average age of software developers in this study being 29 is a plus to the survey data as they fall in the late 20s. Those who agreed to participate were asked to read the scenario and understand the Unified Modeling Language (UML) diagrams as the interview questions will be about them. To confirm this, one of the first interview questions to ask participants before starting the main study was whether they have read and understood the scenario and the UML diagrams. More time was given for those who did not understand, and an explanation offered to ensure that the scenario is understood clearly. The table 2 below shows the participants' demographics.

|  | Minimum. | Maximum. | Average. |
|---|---|---|---|
| Age(years) | 24 | 38 | 29 |
| Duration of interview(minutes) | 39 | 56 | 46 |
| Experience(years) | 4 | 11 | 6 |
| **Gender information.** | | | |
| Male | | Female | |
| 17 participants | | 3 participants | |
| The average rate of coding (per week) | | 5 days | |

**Table II: Demographics Information**

The participants were asked questions that enabled us to investigate how they implement each GDPR principle. We asked the participants if they knew any data privacy law. This was to know whether they were familiar with the GDPR principles or law. There was a set of questions aimed at investigating a particular GDPR principle. An example of the question was whether they could inform data subjects of a data breach if it occurred and how long it would take to inform the subjects. This was aimed at investigating the first GDPR principle i.e. Lawfulness, Fairness, and Transparency. Data Minimization was investigated by asking the participants which set of data they would collect from users in the scenario given.

The questions we asked, enabled us to investigate the six principles of GDPR. We asked "why" follow-up questions from the answers provided by the participants to get more information about their understanding of GDPR e.g., when the answer was "I can collect name, username and password", we asked "why collect this"? We wrote down the responses of the developers as the interview progressed.

We applied grounded theory to analyze the participants' descriptive answers [27]. Some of the answers were yes/no and this did not require grounded theory [27] to analyze. From the 22 participants, we were able to analyze 20 participants' responses as the remaining 2 lacked quality. We applied three coding schemes to the data collected [27]. This was important as it enabled us to capture all the issues that developers face. The first coding we applied is open coding [28]. In this coding scheme, we went through the answers several times and then created tentative labels for chunks of data that summarize the responses. The second coding scheme we applied was axial coding [28]. Here, we identified the relationship between the open codes. We also categorized similar answers in this stage. We were able to come up with many categories in this stage. To link these categories into fewer categories that enabled us to analyze the data more accurately, we used selective coding [28]. From this analysis, we were able to identify why developers cannot use GDPR principles when embedding privacy into software systems

## 4. RESULTS AND DISCUSSION

From the coded data, we were able to identify several issues on why developers cannot embed privacy using GDPR principles. The issues are discussed below:

**a) Participants did not have good techniques to implement GDPR principles.**

Our study on the techniques that the participants use to implement different techniques revealed that most developers did not have the techniques to implement GDPR principles effectively. The GDPR principles, which lacked the highest number of techniques, are Storage Limitation, Purpose Limitation, and Accuracy, in that order. Almost all of the participants lacked techniques to implement different principles. Nine, eight, and six of the participants said that they did not have an idea about how to implement Storage Limitation, Purpose Limitation, and Accuracy respectively. For example, one developer said that *"I don't have an idea about how I to implement Storage Limitation and Purpose Limitation, I have never implemented this before"*. We identified this issue as implementing GDPR law (i.e. GDPR principles) partially, is similar to not implementing the law at all [3]. Our study also revealed that participants could not differentiate different techniques for different GDPR principles. We noticed that the principles that brought this confusion to the participants were: Purpose Limitation; and Lawfulness, Fairness, and Transparency. Participant eight said, *"this principle is confusing, especially purpose limitation and the first principle"* (lawfulness, fairness and transparency).

Eight participants identified different techniques on Lawfulness, Fairness, and Transparency but when asked questions about Purpose Limitations, they said they had already answered those questions when responding to Lawfulness, Fairness, and Transparency. Almost all developers had techniques such as encryption and encapsulation to implement Integrity and Confidentiality. The bar graph below shows the five principles and the number of participants who did not have techniques to implement.

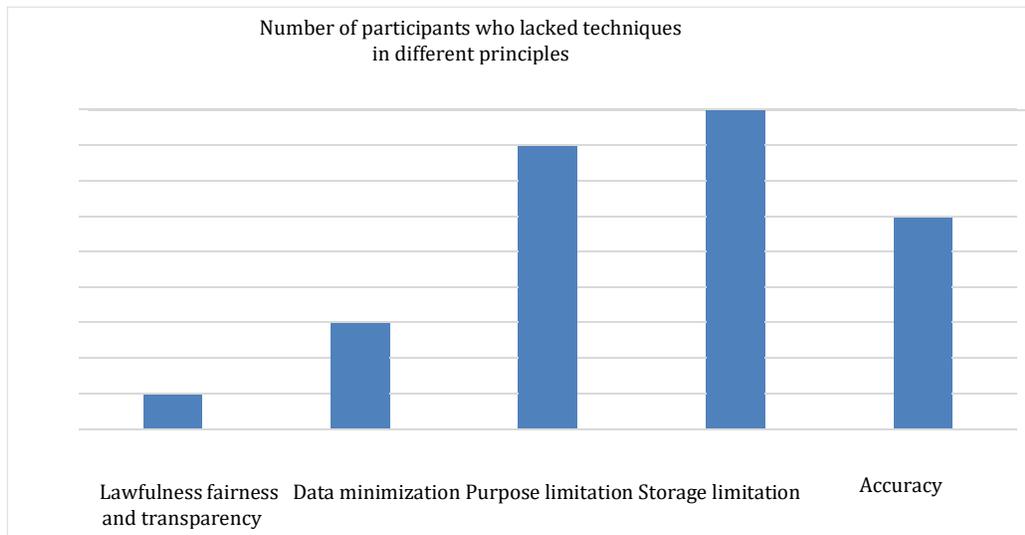

**Fig 6: Number of participants who lacked techniques in different principles**

**b) Participants were not familiar with GDPR principles (i.e., GDPR law).**

Seventeen of the developers we interviewed did not have an idea of the GDPR law. For example, participant ten said that, *"I have never heard of GDPR, but I know data laws in my country"*. Six developers knew about different data privacy laws e.g., one developer said that his country has some data privacy laws, which he uses while developing software. Eleven developers said that they did not have an idea of what data privacy is. Our study revealed that GDPR law is not well known to the software developers. Three developers were familiar with the law but still had difficulties on the implementation of some principles. All the three were from Europe. For example, one developer who was familiar with GDPR said that *"I only know about Integrity and Confidentiality principle"*. Our study found this to be the main problem as people cannot implement concepts or principles (e.g., GDPR principles) they are not familiar with. The bar graph indicates the familiarity of participants with data laws.

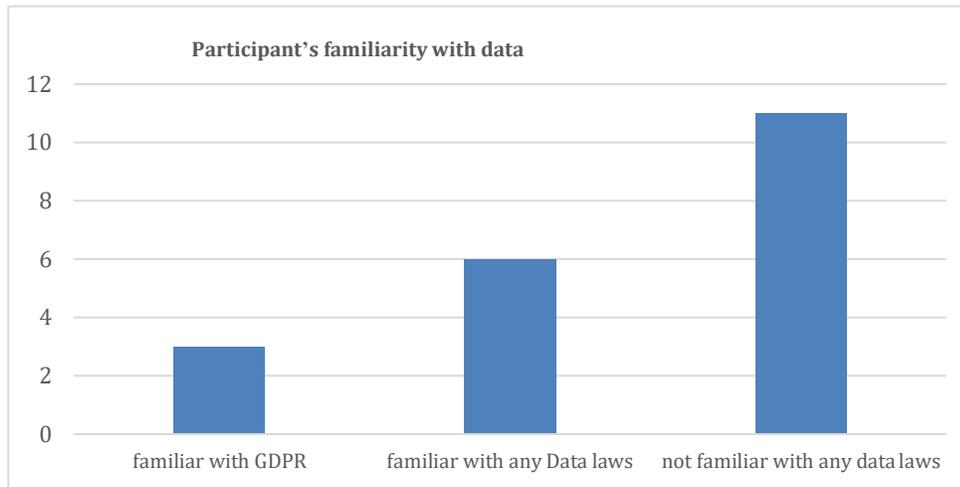

**Fig 7: Participants' familiarity with data laws.**

**c) Participants were more concerned with the functional requirement.**

Nine of the participants claimed that they were more concerned with the functionality of the scenario. For example, one developer said that *"when a customer wants me to develop software for her, I am much more concerned with the functionality of the system. Security and privacy laws are important, but they are not a primary need"*. Our study also revealed that most of the participants believed that implementing software data privacy is not their role. Ten developers claimed that they do not care about software privacy as there is supposed to be a separate team whose sole responsibility is to ensure that software privacy is observed. One participant said that *"for the privacy to be implemented effectively, there should be a team whose role is software privacy development. My main role is to implement customer requirements"*. We also revealed that, about data privacy, they were only concerned with Integrity and Confidentiality. For example, one of the participants said that, *"integrity and confidentiality are the key to privacy, the others are not crucial"*

**d) Participants lacked resources and online materials for reference and guidance when implementing data privacy.**

Our study revealed that the majority of the participants had an issue with the resources and online materials that could guide them to implement data privacy. This issue was identified in almost all the GDPR principles. For example, one of the participants said that the data laws state what should be done but do not state how it should be done. The participants gave an example of GDPR law that has six principles (i.e., GDPR principles) but does not state which techniques developers should use or how they should be followed when implementing the principles in the software. Our study also revealed that this issue leads to the developers using techniques, which are outdated hence not effectively implementing data privacy. One of the answers from the participants stated, *"due to the lack of guidelines, I end up using substandard techniques to implement the data privacy"*. He claimed that for each principle e.g., confidentiality, a list of techniques such as '*use of biometrics or multifactor verification to implement that and privacy library made available online*' should be listed. Storage limitation and purpose limitation were the leading principles, which lacked online resources on how to implement. The bar graph below represents those who claimed they lacked guidelines and online resources to refer to, in each principle.

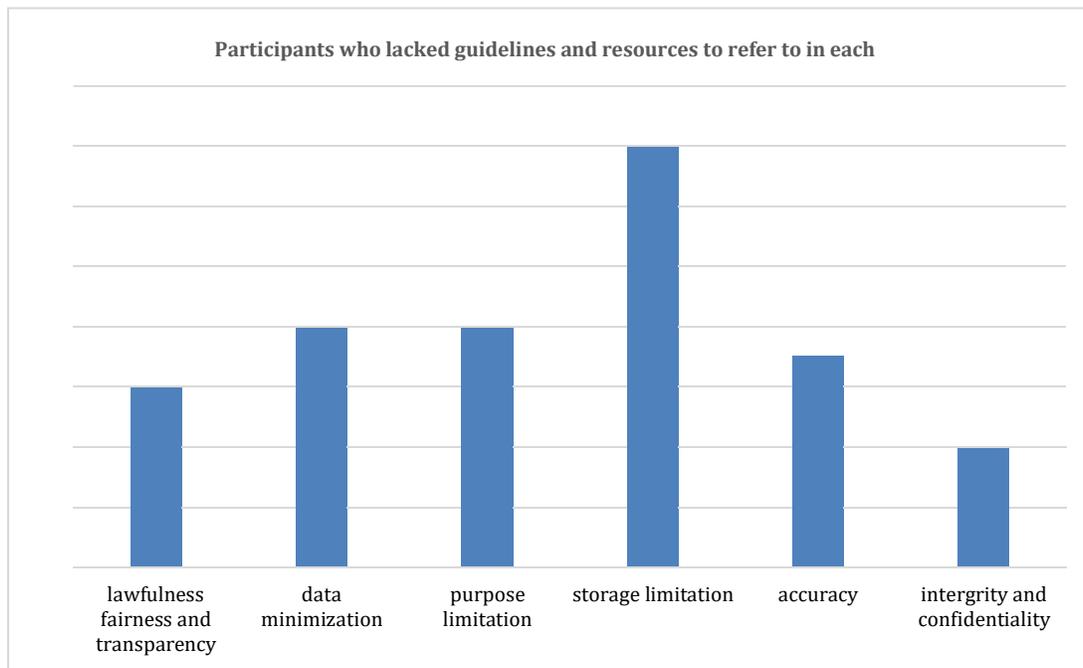

**Fig 8: Participants who lacked guidelines and resources to refer to in each principle**

e) **The implementation of privacy also depends on the client (organizations or the customer).**

Our investigation revealed that organizations play a critical role in ensuring that developers to implement all the GDPR principles effectively. One participant said, "*I may have an idea of all these principles, but if the organizations do not like them, I cannot implement them*". Another participant said that the most efficient methods to effectively implement GDPR principles e.g., multifactor authentication, are expensive to implement, and organizations may choose not to use these methods. Our study revealed that for developers to use GDPR principles effectively when developing privacy-preserving software systems, organizations should be supportive. Eight developers said that organizations affect how effectively they implement the GDPR law.

The five issues, which represent the main findings of our study, are shown in table 3 below

| Issues | Representative Quotes | Number of responses |
|---|---|---|
| Lack of familiarity with GDPR | I am not familiar with GDPR | 17 |
| Organizations | To use GDPR effectively depends on organization you're working for. | 8 |
| Lack of resources and guidelines | Don't have guidelines on how to do it and where to refer to | 16 |
| Not familiar with techniques | I don't have an idea of technique to do that | 14 |
| Privacy is not a primary requirement | My greatest concern is functional requirements. | 9 |

**Table III: Main issues identified**

5. RECOMMENDATIONS

We recommend the following guidelines to address the issues we identified in our investigation:

- GDPR law (or an equivalent standard) should be made an international law; all nations can adopt it as this will enable developers to familiarize themselves with the law and incorporate it into the software systems that they develop for clients as our study revealed that most of developers outside Europe were not familiar with GDPR.
- GDPR law should be accompanied by techniques to use in each principle in order to implement the law effectively and avoid developers using substandard techniques.
- Tutorials on different techniques that aid in the implementation of GDPR principles in software should be made available online. Different resources such as security/privacy libraries should also be readily available to developers.
- Fines and penalties should be imposed to organizations or/and individuals who fail to implement data privacy in software system taking GDPR on board.
- Developers and organizations should be given formal knowledge on the importance of implementing GDPR principles in their software systems and the consequences of failing to comply with GDPR law. Developers should be given formal knowledge (e.g., education and training) on the importance of privacy in software.

These recommendations will help to address the issues that we identified in our study. Our study revealed that these were the main issues, which were common to all the participants. Our study contributes to improving the privacy of personal data taking GDPR on board when implementing privacy-preserving software systems. If developers complied with GDPR law, data breaches may significantly reduce.

6. **Discussion.**

GDPR advocates the introduction of data privacy considerations into the software system. Therefore, software developers responsible for creating software systems plays a significant part in ensuring that GDPR is adhered to. Therefore, it is essential to understand developers' challenges when they implement privacy taking GDPR on board. This study provides qualitative research that attempts to find what challenges developers face when implementing privacy into software systems. This research investigated each principle of GDPR. Researchers have previously conducted investigations on why developers find it challenging to implement GDPR when developing privacy-preserving software systems. Most of them focused on one principle of GDPR law. According to [5], who did a survey to evaluate what hinders developers when implementing privacy, software developers believed that privacy requirements contradict systems requirements. They also found that developers lacked knowledge of privacy practices. Although their research involved one principle of GDPR, i.e., data minimization, these findings are similar to the current study findings. In addition, the current study went the extra mile on investigating all the principles, which enabled us to get more concrete findings, which led to better recommendations.

Our findings indicated that organization impacted how developers implemented privacy taking GDPR. Some researchers, for example [21], who were more concerned with organization practices that affect developer's implementation of privacy, found out that organization practices are important tools that can affect developers' implementation of privacy. This has been seconded in this research as some of the participants claimed that organizations affect how they implement privacy. In this research, we found out that not only the environment but also a specific client as different clients have different requirements and specifications such as budget, which may affect the implementation of privacy observing the GDPR law.

We also found out developers lacked awareness, implementation techniques, tools and reference sources/materials that can guide them when implementing data privacy taking GDPR on board. This was also revealed by [18] in his report. These findings were more detailed as we were able to find out many participants lacked implementation techniques, tools and reference resources online in each element of GDPR. Thus, we found out some elements such as purpose limitation are under-looked. We were more detailed in lack of awareness as we discovered that this was the case with developers outside Europe. This led us to recommend that GDPR law (or an equivalent standard) should be made an international law.

Lastly, we found out that participants were more concerned with functional requirements, which was also revealed by [5]. We found from the research that the majority perceived that privacy involved integrity and confidentiality only, which led us in recommending that formal knowledge on privacy should be given to developers. Based on this finding, it is clear that developers find difficulties when implementing privacy using available regulations, as evident from other researchers. By investigating developers from different localities, we found that developers outside Europe had more challenges when it comes to GDRP implementation. The contributions in this research include empirically investigating challenges faced by developers when implementing privacy through GDPR law and analysis of data that led to investigating each principle separately. These might be reasons data breaches still occur in software systems.

## 7. Limitation.

Some limitation should be considered in this study. One of the limitations is the sample size. It is true that privacy implementation is multifaceted, i.e., it depends not only on software developers but also on other parties e.g., organizational settings, culture, and the nature of the software project/application they develop etc. We have only considered the healthcare application as a scenario reported in this study due to its easiness to depict privacy related issues.

Although this study followed qualitative approach, increasing the sample size could have captured more issues and also achieved higher diversity. Another limitation is that our research was conducted virtually due to the existence of a global pandemic COVID-19 at the time of study, hence the results would have been biased as some participants may have given biased data. Participants (i.e., software developers) responses might be biased due to the challenging environment created by the COVID-19 pandemic. For example, software developers' responses may be biased if there is an uncertainty of losing their jobs. Finally, caution should be taken when generalizing our results due to our limited sample. Future research may involve large number of participants thus allowing quantitative analysis, for example, survey studies.

## 8. CONCLUSION AND FUTURE WORK

This research attempts to identify the issues that prevent software developers from embedding privacy into software, taking GDPR principles (i.e., GDPR law) on board. Our findings enabled us to derive recommendations that would effectively support developers when they embed privacy (i.e., when developing privacy-preserving software systems) to take GDPR law on board.

This research revealed that:
- Participants lacked resources and online materials for reference and guidance when implementing data privacy.
- The implementation of GDPR, when developing privacy-preserving software systems, is affected by organizations or the customer.
- Participants were more concerned with the functional requirement.
- Participants were not familiar with GDPR law or principles.
- Participants did not have good techniques to implement GDPR principles (i.e., GDPR law).

In the future, a study involving more participants may be carried out, to be able to generalize our results. We will also create a game-based education model in future to enhance developers' implementation of privacy in software systems.

**APPENDIX**

**Invitation message**

Dear participant,
 I am a researcher in computer science, doing PhD at XYZ University. My supervisor is Dr. XXX. I am looking for software developers who will participate in a research about software development and I would like to invite you to be a participant of this study. This research targets software developers hence you being one of them, you're the most preferred in this study. This study is composed of a cognitive walk through after reading and understanding a given scenario. The study will not take much of your time, and it will, at most, take 60minutes. We would like to know if you're willing to be a participant in this study. Being in this study is voluntary. Please reply to this message if you're willing to be a participant.

**SCENARIO**

Study the scenario given, after this a cognitive walk through will follow.
*You are required to structure an online health application (web application) that permits patients to register/login and discuss remotely with clinical experts for pay. Clients should be in a position to communicate with the clinical experts in this application on their issues. The clinical expert can join this system to earn from it. The application will get a commission from each paid service by the clients*

**The guiding questions.**

1. What is your experience as a software developer?
2. Do you mind giving me your age?
3. What is your nationality?
4. How often do you code?
5. Have you read the scenario and understand it? If the participant says yes, continue. If the participant says no, explain to the participant
6. If you have read and understood the scenario, I will ask you some questions about the scenario, do I have your permission to go on and use your data in the analysis?
7. What comes to your mind when you hear about privacy in software systems?
    a. Apart from cyber-attacks, do you think there are other elements of software privacy?
    b. As a developer, what do you think is your role in ensuring the privacy of software systems?
    c. Are you familiar with any data protection law? If so, can you please describe them in detail, including how you

      would implement them in the above scenario? (If the answer is no, ask what he uses instead)
    d. What are the difficulties that you will face when implementing the above healthcare scenario that complies with the data laws/ ways you use?
    e. What would you require to curb these challenges?
8. In this scenario, it is required that data be protected from bad-guys (i.e., cybercriminals), as this can lead to illegal access to patients' health record data through the application. The application also has a functional requirement that allows different users to communicate remotely. It will be required that security measures are applied to protect data to avoid data getting into the wrong hands. Can you please explain how you would achieve these functions in the health application and the techniques you will use in the health application?
    a. Why do you think these techniques are best?
    b. Have you ever used these techniques before?
    c. Do you find any difficulties when implementing this?
    d. To curb these challenges, what support would you require?

9. In this health application, the data storage security measures will also be implemented in this application to ensure that unauthorized personals do not access patients' health record data. Can you please explain how you would achieve these functions in the health application?
    a. Why do you think these techniques are the best?
    b. Can you please talk about how would you implement these features into the healthcare application/what technologies or tools you would use?
    c. Do you know any techniques other than these?
    d. (Ask developers the techniques that are the best if he doesn't mention them, e.g. two-way verification and use of biometrics)
    e. Have you experienced (technical or non-technical – e.g., issues with documentation or lack of resources on the internet or YouTube) any problems when implementing the above techniques, you mentioned?
10. In this scenario, the doctors, nurses and patients' data will be collected. Can you please explain how you would develop the healthcare application in order to process the collected data lawfully and fairly?
    a. What are the difficulties when implementing the processing of collected data lawfully and fairly?
    b. Can you please explain how the users are informed in case of getting their data compromised by hackers or cybercriminals?
    c. What is the process that you will inform them or is there any technique you would use to inform users
    d. Would you tell the data subjects all the reasons for collecting their data?
11. When implementing the above healthcare application scenario, what data would you collect from patients, nurses, and doctors?
    a. Please let me know the reason for collecting each the above data.
    b. Do you think it is important to minimize the data we collect from patients, nurses, and doctors through the developed healthcare application? Can you please explain why?
    c. Can you please explain how you would develop the healthcare application to collect minimum data from patients, nurses and doctors?
    d. What challenges would you face when implementing this? And do you lack the guiding resources?
12. In this health application, it will be a requirement to ensure that patients, nurses and doctors know the reason for collecting their data. It will also be required that this data is not to be used for any other purposes than the stated ones. Can you please explain how you would achieve this when developing the above health care scenario application?
    a. How would you inform them about the purpose of collecting their data in the health care application? What techniques do you use?
    b. Have you experienced any difficulties when implementing this? Can you please describe them in detail (i.e., lacking any technical or tooling support, lacking any online resources to refer, e.g., YouTube?)
    c. Have you implemented this before? (If no, ask participant why not and get deeper details, if yes ask how he implemented and challenges he faces)
13. In this health scenario, there are functional requirements such as login, registration, and payment process. Would you consider collecting all the personal data of patients, nurses and doctors when implementing the health care application?
    a. How would you identify (i.e., what techniques or processes) only the required data to be collected when developing the healthcare application?
    b. Can you please explain your experience of developing this feature before?

c. Have you experienced any difficulties when implementing this?
   d. Do you lack any tools/software or online resources (YouTube or online materials) to refer? (In details).
14. Another requirement will be to approve the registration of users. It will also be a requirement in this application to prevent unauthorized persons from accessing user data. Can you please explain how you would achieve these functions when developing in the health application?
   a. Can you please explain your previous experience of implementing this before if you have? (User registration approval)
   b. Have you experienced any difficulties when implementing registration approval? Do you get resources to refer to when implementing this, e.g., on YouTube?
15. The health application will require that the stored records not to be held by the system for a long period once they are no longer needed. This will help in case of data breaches as minimum data leaks to the unwanted personal. Can you please explain how you would implement storage limitations in the health application?
   a. (If the answer is yes asking how they implement, which techniques they use and the challenges they face, if no give hints of techniques, e.g. strategic deletions, and ask developers if they are familiar with them)
   b. In